\documentclass[aps,prb,showpacs,twocolumn,floats]{revtex4}
\usepackage{amssymb}
\usepackage{amsbsy}

\usepackage{amsmath}
\usepackage{epsfig}

\begin{document}

\title {Bose-Fermi mixtures in an optical lattice}

\author{K. Sengupta$^{(1)}$, N. Dupuis $^{(2,3,4)}$, and P. Majumdar $^{(5,6)}$}
\affiliation{$^{(1)}$ Theoretical Condensed Matter Physics Division
and Center for Applied Mathematics and Computational Science \\ Saha
Institute of Nuclear Physics, 1/AF Bidhannagar, Kolkata-700064,
India. \\ $^{(2)}$Department of
Mathematics, Imperial College, 180 Queen's Gate, London SW7 2AZ, UK. \\
$^{(3)}$Laboratoire de Physique des Solides, Univ. Paris-Sud, CNRS,
UMR 8502, F-91405 Orsay Cedex, France.\\ $^{(4)}$  LPTMC, CNRS-UMR
7600, Universit\'e Pierre et Marie Curie, 75252 Paris C\'edex 05,
France. \\ $^{(5)}$ Cavendish Laboratory, Cambridge University,
Madingley Road, Cambridge, CB3 0HE, UK. \\ $^{(6)}$ Harish-Chandra
Research Institute, Chhatnag Road, Jhunsi, Allahabad 211019, India.
}

\begin{abstract}

We study an atomic Bose-Fermi mixture with unpolarized fermions in
an optical lattice. We obtain the Mott ground states of such a
system in the limit of deep optical lattice and discuss the effect
of quantum fluctuations on these states. We also study the
superfluid-insulator transitions of bosons and metal-insulator
transition of fermions in such a mixture within a slave-rotor
mean-field approximation, and obtain the corresponding phase
diagram. We discuss experimental implications of our results.

\end{abstract}


\maketitle

\date{\today}

\section {Introduction}

Recent experiments on ultracold trapped atomic gases have opened a
new window onto the phases of quantum matter \cite{Greiner1}. A gas
of bosonic atoms in an optical or magnetic trap has been reversibly
tuned  between superfluid (SF) and insulating ground states by
varying the strength of a periodic potential produced by standing
optical waves \cite{Greiner1,Orzel1}. This transition has been
explained on the basis of the Bose-Hubbard model with on-site
repulsive interactions and hopping between nearest neighboring sites
of the lattice. \cite{Fisher89, Jaksch1, sesh1, van1, Sengupta1}.
Further, theoretical studies of bosonic atoms with spin and/or
pseudospin have also been undertaken
\cite{imambekov1,demler1,prokofiev1,issacson1}. These studies have
revealed a variety of interesting Mott phases and
superfluid-insulating transitions in these systems. On the fermionic
side, the experimental studies have mainly concentrated on the
observation of paired superfluid states \cite{Greiner2} and the
BCS-BEC crossover in such systems near a Feshbach resonance
\cite{Zwierlein1}.

More recently, it has been possible to generate mixtures of
fermionic and bosonic atoms in a trap \cite{exprev1, modugno1}.
Initially, the main focus of such experimental studies were to
generate quantum degenerate Fermi gases, through sympathetic cooling
with bosons. However, a host of theoretical studies followed soon,
which established such Bose-Fermi mixtures to be interesting
physical systems in their own right
\cite{roth1,albus1,lewenstein1,cramer1,yu1}, exhibiting exciting
Mott phases in the presence of an optical lattice. In all of these
works, the spin of the fermions in the mixture is taken to be frozen
out due to the presence of the magnetic trap. However, more recent
works reported in Refs.\ \onlinecite{illu1} and \onlinecite{carr1},
considered a Bose-Fermi mixture in an optical trap, where the spins
of the fermions can be dynamical degrees of freedom \cite{comment0}.
It has been shown in Ref.\ \onlinecite{illu1} that the interaction
between the bosons and the fermions in such a mixture can enhance
the s-wave pairing instability of the fermions.

In this work, we consider a Bose-Fermi mixture in an optical trap
and in the presence of an optical lattice and study the Mott phases
and the metal/superfluid-insulator transition for the
fermions/bosons of such a mixture using a slave-rotor mean-field
theory \cite{florens1}. The motivation for such a study is two-fold.
First, it has been shown in Ref.\ \onlinecite{lewenstein1} that the
Mott phases of Bose-Fermi mixtures in a magnetic trap are
interesting in their own right. In the present study, we chart out
the Mott phases of the Bose-Fermi mixture in an optical trap, where
the Fermion spins are dynamical degrees of freedom, in the deep
lattice limit for a wide range of parameters. As expected, we find
that the corresponding Mott phases obtained are much richer than
their counterparts studied in Ref.\ \onlinecite{lewenstein1}.
Second, the metal/superfluid-insulator transition of the
fermions/bosons in such an interacting Bose-Fermi mixture has not
been studied before. Here we develop a self-consistent slave-rotor
mean-field theory to study such a transition for the Mott phases
which do not have density-wave order with broken translational
symmetry and use it to obtain at least a qualitative understanding
of the effect of interaction between the fermions and the bosons on
the metal/superfluid-insulator transition.

In what follows, we shall assume that the atoms are confined using
an optical trap so that Fermions spins are not frozen out. We shall,
however, ignore the effect of the harmonic trap potential which is a
standard approximation used extensively in the literature
\cite{florian1}. The starting point of our study is the Bose-Fermi
Hubbard Hamiltonian that has been developed earlier, with similar
approximation regarding the trap potential, from underlying
microscopic dynamics of the atoms in the presence of an optical
lattice \cite{albus1}
\begin{eqnarray}
{\mathcal H} &=& {\mathcal H}_F + {\mathcal H}_B + {\mathcal H}_{FB}
\label{hham} \\
{\mathcal H}_F &=& -t_F \sum_{\left<ij\right> \sigma}
\left(c_{i\sigma}^{\dagger} c_{j\sigma} +{\rm h.c}\right)
\nonumber\\
&& -\mu_F \sum_{i\sigma} n_{i\sigma}^F  + U_{FF} \sum_i
n_{i\uparrow}^F n_{i\downarrow}^F
\label{fermiham}\\
{\mathcal H}_B &=& -t_B \sum_{\left<ij\right> \sigma}
\left(b_{i}^{\dagger} b_{j}+{\rm h.c}\right)\nonumber\\
 && -\mu_B\sum_{i}
n_{i}^B + \frac{U_{BB}}{2} \sum_i n_{i}^B \left(n_{i}^B -1\right)
\label{boseham} \\
{\mathcal H}_{FB} &=& U_{FB} \sum_{i\sigma} n_{i\sigma}^F n_i^B
\label{fbham}
\end{eqnarray}
Here $c_{i\sigma}$ is the fermionic destruction operator with spin
$\sigma= \uparrow, \downarrow$ at site $i$, $b_i$ represents bosonic
destruction operator at site i, $n^{F(B)}$ denotes Fermion(Boson)
number operators, $t_{F(B)}$ and $\mu_{F(B)}$ are nearest neighbor
hopping matrix elements and chemical potentials for the fermions
(bosons), $U_{BB}$ and $U_{FF}$ are the on-site Hubbard repulsion
for bosons and fermions respectively, and $U_{FB}$ denotes the
relative interaction strength between the bosons and the fermions.
In what follows, we shall take the bosons and the fermions to have
fixed chemical potentials $\mu_{B(F)}$ and same on-site repulsion
$U_{BB} = U_{FF} =U $ and consider $\lambda = U_{FB}/U$ and $\eta=
t_F/t_B$ as parameters which can be freely varied. The justification
of this choice is briefly outlined in Sec.\ \ref{mottsec}. Further,
we shall only deal with case of a square bipartite lattice in this
work since this is simplest to realize experimentally.

The organization of the rest of the paper is as follows. In the next
section, we identify the Mott phases of Eq.\ \ref{hham}. Next, in
Sec.\ \ref{mft}, we introduce the slave rotor formalism and use it
within a mean-field approximation to study the
metal/superfluid-insulator transition of the Bose-Fermi Hubbard
model (Eq.\ \ref{hham}). This is followed by a discussion of
possible experiments in Sec.\ \ref{expt}. A comparison of the
Mott-Hubbard phase diagram obtained using the projection operator
technique with those obtained from mean-field theories
\cite{Fisher89,van1,sesh1} and standard strong coupling expansions
\cite{freericks1} is given in Appendix \ref{appa}.

\section {Mott Phases}
\label{mottsec}

In this section, we chart out the Mott phases of the Bose-Fermi
system. To do this, we first obtain the phases of the system in the
Mott limit ($t_B=t_F =0$) and then obtain fluctuation corrections
over these states to ${\rm O}(t_{B(F)}^2/U^2)$.

Before obtaining the Mott phases for the Bose-Fermi mixture, let us
look briefly into the parameters of the Hubbard model (Eq.\
\ref{hham}). These can be determined from the microscopic quantities
such as the potential depths $V_{F(B)}$ due to the laser seen by the
atoms and their recoil energies $E^R_{F(B)}= \hbar^2
k_L^2/2m_{F(B)}$ where $k_L$ is the wave-vector of the laser and
$m_{F(B)}$ are the masses of the fermions(bosons). The potential
depth seen by the atoms depend on the detuning of the laser from
their natural wavelengths $\lambda_{F(B)}$ of the fermions(bosons).
In fact, it can be shown that, the ratio of the lattice potentials
seen by the fermions and bosons are \cite{illu1}
\begin{eqnarray}
\frac{V_F}{V_B} &=& \frac{\lambda_F^4 \Gamma_F \Delta
\lambda_B}{\lambda_B^4 \Gamma_B \Delta \lambda_F} \label{detuning}
\end{eqnarray}
where $\Delta \lambda_{F(B)}= \lambda_L -\lambda_{F(B)}$ denote the
detunings for the fermions(bosons) and $\Gamma_{F(B)}$ are the
corresponding natural linewidths. Since the ratio of the natural
linewidths is generally close to unity \cite{illu1,exprev1}, we see
that one can tune the ratio of the lattice depths seen by bosons and
fermions by varying the detunings.

In terms of these quantities, we have \cite{albus1}
\begin{eqnarray}
t_{B(F)} &=& \left(2/\sqrt{\pi}\right) \left(E^R_{B(F)} V_{B(F)}^3
\right)^{1/4} e^{-2\sqrt{\frac{V_{B(F)}}{E^R_{B(F)}}}} \nonumber\\
U_{BB(FF)} &=& \sqrt{8/\pi} \left(E^R_{B(F)}V_{B(F)}^3 \right)^{1/4}
\, k_L a_{BB(FF)} \nonumber\\
U_{FB} &=& \frac{\left(E^R_{F}V_{B}^3V_F^3 \right)^{1/4}(1+m_F/m_B)
k_L a_{BF} }{\sqrt{\pi/16}\left(
\sqrt{V_B} + \sqrt{V_F E_B/E_F} \right)^{3/2}} \nonumber\\
\label{param}
\end{eqnarray}
where $a_{FF}$, $a_{BB}$, and $a_{FB}$ are the s-wave scattering
lengths for interaction between two fermions, two bosons and a
Fermion and a Boson respectively. These scattering lengths also can
be varied either by choosing different species of fermions or bosons
or by tuning them using Feshbach resonance. Further, as we have
discussed before, by choosing the laser detuning we can also make
the fermions and bosons see either similar or very different lattice
potentials. Therefore, instead of calculating these parameters from
the microscopics, we shall aim to portray a general picture of the
Mott phase diagram. Since the experimental possibilities are
limitless, for the sake of brevity, we choose $U_{BB}= U_{FF}=U$ and
vary the ratios $\lambda = U_{BF}/U$ and $\eta= t_F/t_B$. It is
clear from the above discussions that such a situation can be always
achieved in experiments. We shall consider some such specific
examples in Sec.\ \ref{expt}.

Next, we consider the Bose-Fermi Hubbard Hamiltonian in the Mott
limit. In this limit, the on-site states can be represented as
$\left|n_0^B,n_0^F\right>$ and the energy is given by
\begin{eqnarray}
E\left[n_0^B,n_0^F\right] &=& E_F[n_0^F] + E_B[n_0^B] +
E_{FB}[n_0^B,n_0^F]
\nonumber\\
E_F[n_0^F] &=& -\left(\mu'_F -\frac{1}{2}\right) n_0^F + \frac{1}{2}
\left(n_0^F-1\right)^2 \nonumber\\
E_B[n_0^B] &=& -\mu'_B  n_0^B +\frac{1}{2} n_0^B \left(n_0^B -1
\right) \nonumber\\
E_{FB}[n_0^B,n_0^F] &=& \lambda n_0^F n_0^B \label{energy}
\end{eqnarray}
where we have scaled all energies by $U$ and $\mu'_{F(B)}=
\mu_{F(B)}/U$. It can be seen from Eqs.\ \ref{energy}, that two
states $\left|n_0^B,n_0^F-1\right>$ and $\left|n_0^B-1,
n_0^F\right>$ are degenerate when
\begin{eqnarray}
(\lambda-1) (n_0^F -n_0^B)=\mu_F-\mu_B \label{deg1}
\end{eqnarray}
whereas three states $\left|n_0^B,n_0^F\right>$, $|n_0^B,n_0^F-1>$
and $|n_0^B-1,n_0^F>$ are degenerate when
\begin{eqnarray}
\mu'_F &=& \left(n_0^F+n_0^B \lambda -1\right) \nonumber\\
\mu'_B &=& \left(n_0^F\lambda + n_0^B  -1\right) \label{deg2}.
\end{eqnarray}
The conditions of these degeneracies, of course, depend on our
choice of parameters of the model. It is also to be noted that in
Eqs.\ \ref{deg1} and \ref{deg2}, $n_0^F$ and $n_0^B$ in the ground
state are themselves functions of $\mu'_{F}$, $\mu'_B$ and
$\lambda$, and have to be determined by minimizing Eq.\ \ref{energy}
subject to the constraint of $n_0^F$ and $n_0^B$ being integers.

\begin{figure}
\centerline{\psfig{file=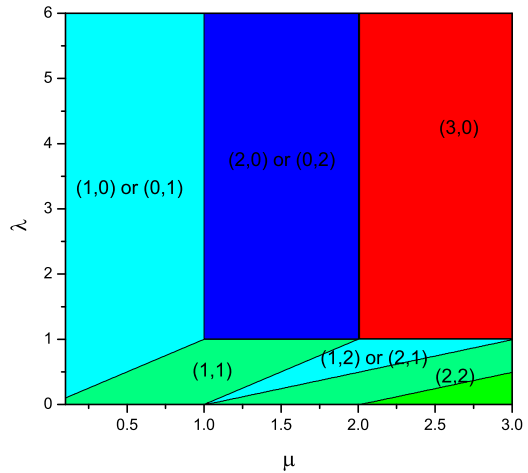,width=\linewidth,angle=0}}
\caption{(Color online) Ground state phase diagram in the atomic
limit for $\mu=\mu'_B=\mu'_F$ and $\lambda>0$. The phases are marked
by values of $(n^B_0,n^F_0)$ in the ground state. For large
$\lambda$, the system tries to avoid putting bosons and fermions at
the same site. The threefold degeneracy between $(1,0)$ and $(0,1)$
as well as $(2,1)$ and $(1,2)$ occur for a large portion of the
phase diagram. In addition, there are doubly degenerate regions such
as $(2,0)$ and $(0,2)$. These degeneracies are lifted by virtual
hopping process for small but finite $t_B$ and $t_F$} \label{fig1}
\end{figure}

The ground state phase in the Mott limit ($t_b=t_F=0$) diagram can
be obtained by numerically minimizing the ground state energy (Eq.\
\ref{energy}) for integers $0\le n_0^F \le 2$ and $n_0^B$. For the
sake of brevity, we carry out the numerical computation for $\mu'_B
= \mu'_F =\mu$ and present the phase diagram as a function of $\mu$
and $\lambda$. The phase diagram for $\lambda>0$ is shown in Fig.\
\ref{fig1}. We find, as expected from the results of Ref.\
\onlinecite {illu1}, for $\lambda > 1$, the fermions and the bosons
repel each other out from a given site so that a site is occupied by
either a Boson or a Fermion, but not both. Such states were dubbed
as "composite" states of bosons/fermions with a correlation hole of
fermions/bosons in Ref.\ \onlinecite{illu1}. However, in the present
scenario, the spins of the fermions are dynamical degrees of freedom
which leads to richer variety of possible phases, as we discuss
below.
\noindent

\begin{figure}
\centerline{\psfig{file=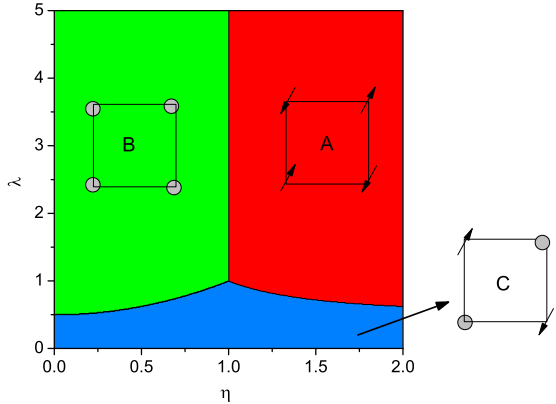,width=\linewidth,angle=0}}
\caption{(Color online) Possible ground states A, B and C that
results from lifting the classical degeneracy of the ground states
$(n_0^B,n_0^F)= (1,0)$ and $(0,1)$ due to quantum fluctuations.
Notice that the antiferromagnetic ordering of state C is different
from state A and is a consequence of very weak next nearest neighbor
hopping. Similar states will result when the degeneracy of the
states $(2,1)$ and $(1,2)$ is lifted. } \label{fig3}
\end{figure}

As expected from discussions leading to Eq.\ \ref{deg1}, a large
portion of phase diagram has degenerate ground states corresponding
to $\left|n_0^B=0,n_0^F=1\right>$ and
$\left|n_0^B=1,n_0^F=0\right>$. Note that in terms of the original
fermions, the degeneracy is actually threefold corresponding to
states $\left|0,\uparrow\right>$, $\left|0,\downarrow \right>$ and
$\left|1,0\right>$. This degeneracy is lifted by quantum
fluctuations due to the presence of small but finite $t_F$ and
$t_B$. This can lead to three different ground states as sketched in
Fig.\ \ref{fig3}: A) an antiferromagnetic state of fermions with no
bosons, B) A state of one boson per site and no fermions, and C) A
state with fermions and bosons being the nearest neighbors with an
antiferromagnetic order for the fermions. The energies of these
states can be estimated using a straightforward ${\rm
O}(t_{B(F)}^2/U^2)$ perturbation theory and are given by
\begin{eqnarray}
E_A &=& -\frac{N z t_F^2}{U} = -\frac{N z t_B^2}{U}\eta^2
\nonumber\\
E_B &=& -\frac{2 N z t_B^2}{U} \nonumber\\
E_C &=& -\frac{N z t_B^2}{2 \lambda U} (1 + \eta^2) \label{qen}
\end{eqnarray}
where $N$ is the total number of sites in the system and $z$ is the
coordination number of each site. Comparing the energies of the
states from Eq.\ \ref{qen}, we find that state A is favored over B
and C when $\eta^2 > 2 $ and $\lambda > (1+\eta^{-2})/2$. Similarly
the state B is favored for $\eta^2 < 2$ and $\lambda > (1+\eta^2)/4$
and state C for small $\lambda$ when $ (1+\eta^2)/(2\lambda) > {\rm
Max}\left[2, \eta^2\right]$. The corresponding phase diagram is
shown in Fig.\ \ref{fig3}.

\noindent

\begin{figure}
\centerline{\includegraphics[width=4cm]{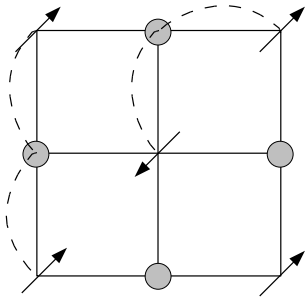}}
\caption{Virtual hopping (dashed lines) leading to frustration of
the
  antiferromagnetic ground state C. This frustration is lifted and an
  antiferromagnetic ground state stabilized by virtual hopping generated by
  the kinetic coupling $t'$ between next-nearest neighbor sites (located at
  opposite corners of the square lattice unit cell).}
\label{fig_frustration}
\end{figure}

With only a nearest-hopping term in the Hamiltonian, the
antiferromagnetic ordering in the ground state C is in fact
frustrated. Indeed, $4th$-order (in $t$) virtual hopping leads to an
antiferromagnetic coupling both between next-nearest-neighbor sites
and next-next-nearest sites (Fig.~\ref{fig_frustration}). In
practice, however, this frustration is suppressed by ${\rm
O}(t'^2/U)$ corrections that arise from the next-nearest neighbor
hopping $t'$ (not included in the Hamiltonian (\ref{fermiham})),
leading to an antiferromagnetic order at wave-vector $(\pi/a,0)$ (or
$(\pi/\sqrt{2}a,\pi/\sqrt{2}a)$ in a reference frame tilted by
$45^o$), in two dimensions as shown in Fig.\ \ref{fig3}. This should
be contrasted with the usual $(\pi/a,\pi/a)$ ordering realized
either for state A or for the (1,1) state shown in Fig.\ \ref{fig1}.
Additional hopping amplitudes (e.g. the next-next-nearest-neighbor
hopping $t''$) are expected to be smaller and will not affect the
antiferromagnetic ordering of the ground state C. A similar
consideration applies for the degenerate states $(2,1)$ and $(1,2)$
where quantum fluctuations will similarly lift the degeneracies.
Notice that the possibilities of having a ferromagnetic state of
fermions where all sites are uniformly occupied by $\uparrow$ or
$\downarrow$ fermions never occur since such a state necessarily
suppresses Fermion hopping due to Pauli principle and is thus higher
in energy compared to the state A.

Finally, we point out that lifting of degeneracy by quantum
fluctuations also occurs for the twofold degenerate states labeled
as $(2,0)$ or $(0,2)$. Note that the state $(2,0)$ corresponding to
two fermions per site requires that all second order virtual hopping
processes are suppressed, unless higher bands are involved. Hence
such a state is energetically unfavorable and is never realized. The
two other states are D) a homogeneous state with two bosons per site
and E) a state with alternate arrangements of two bosons and two
fermions per site. The energies of these states are given by $E_D =
-6Nzt_B^2/U$ and $E_E = -Nzt_B^2(1+\eta^2)/(2\lambda-1)U$. The
latter state (E) is thus favored over state (D) for $\eta^2 > 6
(2\lambda-1)-1$.

\begin{figure}
\centerline{\psfig{file=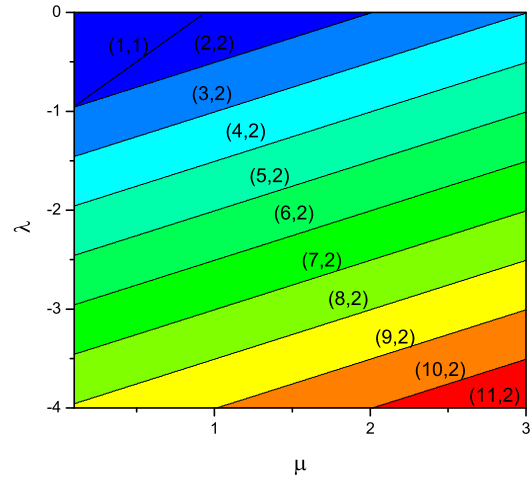,width=\linewidth,angle=0}}
\caption{(Color online) Ground state phase diagram in the Mott limit
for $\lambda <0$. The number of fermions per site is $n^F_0=2$ while
the number of bosons per site is shown in each phase for large
negative $\lambda$.} \label{fig2}
\end{figure}

In contrast for $\lambda <0$ ({\it i.e.} $U_{BF}<0$), there are no
degeneracies. The $(1,1)$ state persists at small negative
$\lambda$. For $|\lambda|>1$, the system always has two fermions per
site with $n_0$ bosons, where $n_0$ is the integer which minimizes
$E[n_0] = -\mu n_0 + n_0(n_0-1)/2 -2 |\lambda| n_0$. The phase
diagram is shown in Fig.\ \ref{fig2} as a function of $\mu$ and
$\lambda$. The phase diagram corresponds to Mott phase of bosons
coupled with non dynamical fixed number of fermions. Note that
analogous composite states with $n_0$ bosons and one Fermion were
found and dubbed as "composite Fermion" states in Ref.\
\onlinecite{illu1}. Here we have almost identical states at large
negative $\lambda$ with the difference that there are two fermions
per site (instead of one per site as in Ref.\ \onlinecite{illu1})
owing to the fact that Fermion spins are not frozen in the present
study.

\noindent

\section{Slave Rotor Mean field theory}
\label{mft}

In this section, we construct a slave-rotor mean-field theory for
the coupled Bose-Fermi problem and use it to study the
metal/superfluid-insulator transition in this system. We develop the
formalism for the mean-field theory in Sec.\ \ref{formalism}, and
discuss the results obtained in Sec.\ \ref{results}

\subsection{Formalism}
\label{formalism}

We begin by implementing the slave rotor formalism \cite{florens1}
in the present context. This key observation behind this formalism
is that the fermionic Hubbard Hamiltonian (Eq.\ \ref{fermiham}) can
be mapped onto a Hamiltonian of free auxiliary Fermions coupled
self-consistently to a quantum rotor. The chief advantage of this
representation is that the quartic interaction term of the original
fermionic Hubbard model can now be represented by a quadratic term
in the rotor variables and can thus be treated exactly in the Mott
limit. This feature makes this technique suitable for studying
Hubbard models in the strong-coupling regime. Further, as shown in
Ref.\ \onlinecite{florens1}, the metal-insulator transition of the
original Fermions can be looked upon as the order-disorder
transition of the rotors which facilitates the study of
metal-insulator transition.

To begin with, we note the identity
\begin{eqnarray}
\frac{1}{2} \left[\sum_{\sigma} \left(n_{i \sigma}^F - \frac{1}{2}
\right) \right]^2&=&  n_{i\uparrow}^F n_{i \downarrow}^F -
\frac{1}{2} \sum_{\sigma} n_{i\sigma}^F +\frac{1}{2} \nonumber\\
\end{eqnarray}
so that, up to a constant term, one can write Eq.\ \ref{fermiham} as
\begin{eqnarray}
{\mathcal H}_F &=& -t_F \sum_{\left<ij\right> \sigma}
\left(c_{i\sigma}^{\dagger} c_{j\sigma} +{\rm h.c}\right)
\nonumber\\
&& -\left(\mu_F-\frac{U}{2}\right) \sum_{i\sigma} n_{i\sigma}^F  +
\frac{U}{2} \sum_i \left[\sum_{\sigma} \left(n_{i \sigma}^F -
\frac{1}{2} \right) \right]^2 \nonumber\\
\label{fermiham2}
\end{eqnarray}
Next, following Ref.\ \onlinecite{florens1}, we introduce the slave
rotor representation for the fermions. The key observation here is
that the spectrum of $H_F$ in the Mott limit depends only on the
total Fermion number $\sum_{\sigma} n_{i \sigma}$ which can be
represented by eigenvalues of angular momenta of a O(2) rotor. Thus
we write the physical fermion annihilation operator as
\begin{eqnarray}
c_{i\sigma} &=& f_{i\sigma} \exp\left(-i \theta_i\right)
\end{eqnarray}
where $f_{i\sigma}$ denotes the annihilation operator for the
pseudo-fermion and $\theta_i$ denotes the rotor variable at site
$i$. The corresponding angular momentum of the rotor is denoted by
$L_i = -i \partial_{\theta_i}$. In this representation, a physical
fermion state can be written as a product of the pseudo-fermion
state and a rotor state as
\begin{eqnarray}
\left|c_1,c_2 ... c_Q \right>_i &=& \left|f_1, f_2, ... f_Q\right>_i
\left| l= Q-1\right>_{\theta_i}.
\end{eqnarray}
Here $l$ denotes the eigenvalues of the rotor angular momentum $L$
and $\left|c_1,c_2 ... c_Q \right> $ is the antisymmetric
combination of Fermionic states for $Q=l+1$ fermions at site $i$.
Note that whereas for ordinary rotors $l$ can take all possible
integer values, here it is constrained to range between $-1$ and $1$
by the operator identity
\begin{eqnarray}
 L_i &=& \sum_{\sigma} \left(f_{i\sigma}^{\dagger}f_{i\sigma} -\frac{1}{2}
 \right) \label{cons}
 \end{eqnarray}
 With the following constraint, one can write Eqs.\ \ref{fermiham2}
 and \ref{fbham} in  terms of the rotor variables as
 \begin{eqnarray}
{\mathcal H}_F &=& -t_F \sum_{\left<ij\right> \sigma}
\left(f_{i\sigma}^{\dagger} f_{j\sigma}
\exp\left[i\left(\theta_i-\theta_j\right)\right]+{\rm h.c}\right)
\nonumber\\
&& -\left(\mu_F-\frac{U}{2}\right) \sum_{i\sigma}
f_{i\sigma}^{\dagger} f_{i\sigma} +
\frac{U}{2} \sum_i L_i^2 \label{fermiham3}\\
{\mathcal H}_{FB} &=& \lambda U\sum_{i\sigma} f_{i\sigma}^{\dagger}
f_{i\sigma} n_i^B \label{fbham2}
\end{eqnarray}
with $L_i$ related to
$\sum_{\sigma}f_{i\sigma}^{\dagger}f_{i\sigma}$ by Eq.\ \ref{cons}.
In the Mott limit($t_F=t_B=0$), we can implement the constraint
(Eq.\ \ref{cons}) exactly and this procedure leads to Eq.\
\ref{energy} with $n_0^F = Q = l+1$ in a straightforward manner.
Note also that the quartic interaction term $U \sum_i n_{i\uparrow}
n_{i \downarrow}$ in Eq.\ \ref{fermiham} has now been replaced by a
quadratic term $U \sum_i L_i^2/2$ in Eq.\ \ref{fermiham3}. This has
been done at the expense of generating a non-linear coupling between
the auxiliary Fermions and the rotors, as is evident from the first
term of Eq.\ \ref{fermiham3}.

When $t_F, t_B \ne 0$, the above mentioned constraint condition can
not be implemented exactly and we need to resort to mean-field
approximation. The slave-rotor mean-field theory for the present
system can be developed by a straightforward generalization of the
formalism developed in Ref.\ \onlinecite{florens1}. To begin with,
we write the system Hamiltonian ${\mathcal H} $ as
\begin{eqnarray}
{\mathcal H} &=& H_r + H_f + H_b \\
H_r &=&  -\sum_{<ij>} \tau_{ij}^{\rm eff}
\cos\left(\theta_i-\theta_j\right)  + \sum_i \frac{U}{2} L_i^2 +
h L_i \label{bh1} \\
H_f &=& -\sum_{<ij>,\sigma} \left(t_{ij}^{\rm eff}
f_{i\sigma}^{\dagger} f_{j\sigma} + {\rm h.c.} \right) \nonumber\\
&& + \sum_{i\sigma} (-\mu_F + U/2 + h + \lambda U {\bar n}_B  )
f_{i\sigma}^{\dagger} f_{i\sigma} \label{bh2} \\
H_b &=& -t_b \sum_{<ij>} (b_i^{\dagger} b_j +{\rm h.c} ) +
\frac{U}{2} \sum_i n_{i}^B ( n_{i}^B-1) \nonumber\\
&& -\left(\mu_b - \lambda U {\bar n}_F \right) \sum_i n_{i}^B
\label{bh}
\end{eqnarray}
where we have implemented the constraint Eq.\ \ref{cons} using an
auxiliary field $h_i$ which has been replaced by it's saddle point
value $h$ at the mean-field level. Here we also treat the coupling
between the bosons and the fermions within mean-field approximation
by replacing the Fermion/bosons density operators $n_i^{B/F}$ by
their averages ${\bar n}_{B/F}$. This amounts to replacing the term
$\sum_i \lambda U n_i^F n_i^B$ by $\sum_i \lambda U n_i^F {\bar n}_B
$ in Eq.\ \ref{bh2}, and by $\sum_i \lambda U n_i^B {\bar n}_F $ in
Eq.\ \ref{bh}. Within this mean-field approximation, the coupling
term acts as a density-dependent shift in the chemical potentials
for the bosons and the fermions. The effective hopping matrix
elements $t_{ij}^{\rm eff}$ and $\tau_{ij}^{\rm eff}$ in Eqs.\
\ref{bh1} and \ref{bh2} are given by
\begin{eqnarray}
t_{ij}^{\rm eff} &=& t_{ij} \left< \cos(\theta_i - \theta_j)
\right>_{H_r} \nonumber\\
\tau_{ij}^{\rm eff} &=& t_{ij} \left< \sum_{\sigma}
f_{i\sigma}^{\dagger} f_{j\sigma} \right>_{H_f}
\end{eqnarray}
with the assumption that averages such as $\left< \exp(\theta_i -
\theta_j) \right>$ and $\left< \sum_{\sigma} f_{i\sigma}^{\dagger}
f_{j\sigma}\right>$ are real on each bond \cite{florens1}.

The next step is to approximate the rotor model by an effective
single site model
\begin{eqnarray}
H'_{r} &=& \sum_i \left (K \cos(\theta_i) + \frac{U}{2} L_i^2 +
h L_i \right) \\
K &=& -2 \sum_j \tau_{ij}^{\rm eff}
\left<\cos(\theta_j)\right>_{H_r}
\end{eqnarray}
Note that this approximation makes sense only when the Mott ground
state does not have density wave order of any kind either for bosons
or fermions. Using the fact that under this approximation the $H'_r$
becomes a single site Hamiltonian, we define
\begin{eqnarray}
Z &=& \left<\cos(\theta)\right>_{H_r}^2 \label{rel3}
\end{eqnarray}
and use $H_f$ to compute all average involving the Fermionic fields.
This yields
\begin{eqnarray}
K &=&  4 \left<\cos(\theta)\right>_{H_r} \int d\epsilon \, \epsilon
\,
n_F \left(Z\epsilon -\mu_F + h + \lambda U {\bar n}_B \right) \nonumber\\
&=& 4 \left<\cos(\theta)\right>_{H_r} {\bar \epsilon} \label{keq} \\
\left<L\right> &=& 2 \left[\int d\epsilon D(\epsilon) \theta
\left(Z\epsilon
-\mu_F +U/2 + h + \lambda U {\bar n}_B \right)\right ] -1 \nonumber\\
&=&  2 {\bar n}_F -1\label{l1eq}
\end{eqnarray}
where the density of states $D(\epsilon)$ is defined as $D(\epsilon)
= \int d^3k \delta ( \epsilon - \epsilon_k ) /(2\pi)^3$, and
$\epsilon_k = -2t \sum_{i=x,y,z}\cos \left(k_i\right)$ is the
kinetic energy of the fermions (all momenta are measured in units of
lattice spacing), and ${\bar \epsilon}$ and ${\bar n}_F$ are the
average fermionic kinetic energy and density respectively. Comparing
the expression of ${\bar n}_F$ with that for the free fermions $
{\bar n}_F = \int^{\mu_0}_{-\infty} d\epsilon D(\epsilon)$, where
$\mu_0$ is the chemical potential for the free fermions at $T=0$,
one has the relation
\begin{eqnarray}
Z\mu_0 &=&  \mu_F -U/2 - h - \lambda U {\bar n}_B\label{rel1} \\
\left<L\right> &=& 2 {\bar n}_F -1 \label{rel2}
\end{eqnarray}
Notice that since $Z$ vanishes at the transition at $\mu_F=\mu_F^c$,
one gets $ \mu_F^c -U/2 = h + \lambda U {\bar n}_B$. Thus $Z$ acts
as the order parameter for the metal-insulator transition of the
fermions.

Eqs. \ref{rel3}, \ref{keq}, \ref{l1eq}, \ref{rel1} and \ref{rel2}
have to be self-consistently solved to obtain the ground state of
the system. However, to do this, one needs to obtain the ground
state of the bosonic Hamiltonian (Eq.\ \ref{bh}) and compute the
average value of the boson density ${\bar n}_B$. Since at the
mean-field level, the average Fermionic density ${\bar n}_F$ enters
the boson Hamiltonian as a shift in the chemical potential, one can
use the projection operator technique developed in Ref.\
\onlinecite{issacson1} for obtaining ${\bar n}_B$. It was shown in
the context of two-species Bosons that the projection operator
method compares well with quantum Monte Carlo results
\cite{issacson1}.

The projection operators for the boson Hamiltonian (Eq.\ \ref{bh})
can be constructed following the procedure of Ref.\
\onlinecite{issacson1}. It is given by
\begin{eqnarray}
P_{l} &=& \left(\left|n_0^B \right>_{i} \left<n_0^B \right|_i
\right) \otimes \left(\left|n_0^B \right>_{j} \left<n_0^B \right|_j
\right)\label{bprojop}
\end{eqnarray}
where $n_0^B$ is the Boson occupation per site for the Mott ground
state and $l$ denotes the link connecting two neighboring sites $i$
and $j$.

The hopping term for the bosons can be rewritten in terms of sum
over links as
\begin{eqnarray}
T &=& \sum_l T_l = -t_b \sum_l \left(b_{i}^{\dagger} b_{j} +{\rm
h.c.} \right)
\end{eqnarray}
where $i$ and $j$ are near neighbor sites connecting the link $l$.
In this notation, one can now divide the hopping terms into two
parts
\begin{eqnarray}
T_l &=& T_l^1 + T_l^0  = \left(P_l T_l + T_l P_l \right) +
P_l^{\perp} T_l P^{\perp}_l \label{bhopping1}
\end{eqnarray}
where $P^{\perp}_l = 1-P_l$.  It is then easy to see that the term
$T^1 = \sum_l T_l^1$ acting on the ground state takes one out of the
ground state manifold. The idea is therefore to seek a canonical
transformation operator $S$ which eliminates $T_1$ to ${\rm
O}(t_B/U)$ from the low energy effective Hamiltonian: $[iS,H_0] =
-T^{1}$, where $H_0$ denotes all the on-site terms in Eq.\
\ref{fbham}. The effective low energy Hamiltonian can be obtained by
the usual Schrieffer-Wolff transformation method
\begin{eqnarray}
H^{\ast} &=& \exp(iS) H \exp(-iS) \\
&=& H_0 + T^{0} + \left[i S,T \right] + \frac{1}{2}
\left[iS,\left[iS,H_0\right]\right] + ... \label{effham1}
\end{eqnarray}
Note that this is equivalent to a systematic $t_B/U$ expansion and
all the omitted terms denoted by ellipsis are at least ${\rm
O}(t_B^3/U^3)$.

The next task is to find out the canonical transformation operator
$S$ in terms of the projection operators $P_l$. Following Ref.\
\onlinecite{issacson1}, we guess the form of the $S$ to be
\begin{eqnarray}
S &=&  i \alpha \sum_l \left[ P_l, T_l\right]
\end{eqnarray}
where the coefficient $\alpha$ is to be determined by the condition
$[iS,H_0] = -T^{1}$. To do this we use the operator identities
\begin{eqnarray}
\left[P_l T_l,H_0 \right] &=& U P_l T_l, \quad  \left[T_l P_l,H_0
\right] = -U  T_l P_l
\end{eqnarray}
and evaluate $[iS,H_0]$ to be
\begin{eqnarray}
[iS,H_0] &=&  - \frac{\alpha}{U} \sum_l \left(P_l T_l + T_l P_l
\right) = -T ^{1} \frac{\alpha}{U}
\end{eqnarray}
Thus we find that setting $\alpha=U$ we obtain the expression for $S
= i\sum_l \left[P_l,T_l\right]/U$, which eliminates $T_1$ to ${\rm
O}(t_B/U)$ from the low energy effective Hamiltonian.

The effective Hamiltonian can be now rewritten by substituting the
condition $[iS,H_0] = -T^1$ in the last term of Eq.\ \ref{effham1}
as
\begin{eqnarray}
H^{\ast} &=& H_0 + T^0 - \sum_{l,l'}\left[\left[P_l,T_l\right],
T^0_{l'} + T^1_{l'}/2\right] \label{effham2}
\end{eqnarray}

With some algebra we now reach the final form of the low energy
effective boson Hamiltonian which takes into account all $t_B^2/U^2$
fluctuations
\begin{eqnarray}
H^{\ast} &=& H_0 + T^0 - \frac{1}{U} \Bigg[ \sum_l \left(P_l T_l^2
P_l - T_l P_l T_l \right) \nonumber\\
&& + \sum_{l,l'} \Bigg\{ P_l T_l T_{l'} - T_l P_l T_{l'} \nonumber\\
&&-\frac{1}{2} \Big( P_lT_lT_{l'} P_{l'} - T_l P_l P_{l'} T_{l'}
\Big) \Bigg \} \Bigg] \label{projham}
\end{eqnarray}
where $l$ and $l'$ are nearest neighbor links. One can now use a
on-site variational wavefunction  in the same way as in Ref.\
\onlinecite{issacson1}
\begin{eqnarray}
\Psi_v &=& \prod_i \left( a \left|n_0^B\right>_i +b
\left|n_0^B+1\right>_i +c\left|n_0^B-1\right>_i
\right)\label{varwave}
\end{eqnarray}
to minimize the ground state energy $E_G = \left<\Psi_v
\right|H^{\ast} \left|\Psi_v \right>$  and obtain the corresponding
boson density ${\bar n}_B$ and superfluid order parameter $\Delta$
\begin{eqnarray}
{\bar n}_B &=& \left|a_G\right|^2 n_0^B + \left|b_G\right|^2
\left(n_0^B+1\right) +
\left|c_G\right|^2 \left(n_0^B-1\right) \label{denf}\\
\Delta &=& \sqrt{n_0^B+1} \, a_G^{\ast} b_G  + \sqrt{n_0^B} \,
c_G^{\ast} a_G \label{ordf}
\end{eqnarray}
where $a_G$, $b_G$ and $c_G$ are values of the coefficients $a$, $b$
and $c$ in the variational ground state. A comparison of the phase
diagram obtained by minimizing $E_v =
\left<\Psi_v\right|H^{\ast}\left| \Psi_v\right>$ for $\lambda=0$
with analogous phase diagrams obtained from mean-field theory
\cite{Fisher89,van1,sesh1} and defect phase calculations to {\rm
O}($t_B^2/U^2$) \cite{freericks1} is presented in App.\ \ref{appa}.
\begin{figure}
\centerline{\psfig{file=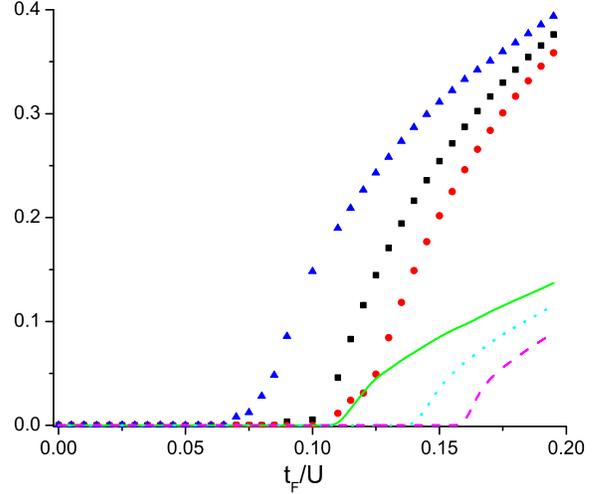,width=\linewidth,angle=0}}
\caption{(Color online) Plot of the order parameter $\Delta$ and $Z$
for $\mu/U=0.7$ and $\eta=5$. Non-zero values of $\Delta$ /$Z$
signals superfluid/metal-insulator transition for the
Boson/fermions. The symbols are as follows: black squares and green
solid line ($\Delta$ and $Z$ respectively for $\lambda=0$), red
circles and cyan dotted line ($\Delta$ and $Z$ respectively for
$\lambda=0.3$), and blue triangles and magenta dashed line ($\Delta$
and $Z$ respectively for $\lambda=0.5$). }\label{fig4}
\end{figure}
\noindent

\begin{figure}
\centerline{\psfig{file=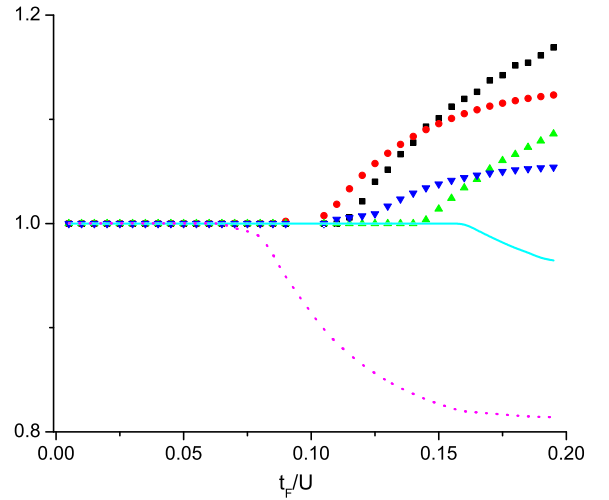,width=\linewidth,angle=0}}
\caption{(Color online) Plot of the average densities ${\bar n}_B$
and ${\bar n}_F$ for the bosons and Fermions for $\mu/U=0.7$ and
$\eta=5$. The deviation of the densities from their quantized values
in the Mott state signals the onset of metal-insulator transition
for fermions and superfluid-insulator transition for the bosons. The
symbols are as follows: black square and red circle (${\bar n}_B$
and ${\bar n}_F$ respectively for $\lambda=0$), green uptriangle and
blue downtriangle (${\bar n}_B$ and ${\bar n}_F$ respectively for
$\lambda=0.3$), magenta dotted line and cyan solid line (${\bar
n}_B$ and ${\bar n}_F$ respectively for $\lambda=0.5$). }
\label{fig5}
\end{figure}
\noindent

Eqs.\ \ref{denf} and \ref{ordf}, combined with Eqs.\ \ref{rel3},
\ref{keq}, \ref{l1eq}, \ref{rel1} and \ref{rel2} can now be solved
self consistently to obtain a mean-field description of the coupled
Fermi-Bose system near the metal/superfluid-insulator transition
points which are signaled by the onset of a non-zero $Z$ or
$\Delta$. This method, therefore, allows for a self-consistent
treatment for the coupled Bose-Fermi problem near the
metal/superfluid-insulator transition point, provided that the
insulating ground state preserves translational symmetry.

\subsection{Results}
\label{results}

In this section, we discuss the results of application of the
formalism developed in the last section to the problem at hand. Here
we shall concentrate on the $(1,1)$ Mott state for the following
reasons. First, since we would be interested in studying
metal-insulator transitions for fermions together with the
superfluid-insulator transition of bosons, we would like to
concentrate on Mott states which has one Fermion per site. Second,
in the formalism developed in Sec.\ \ref{formalism}, we treat the
Bose-Fermi interaction term within mean-field approximation, we
would like to restrict ourselves to the parameter regime
$\left|\lambda\right| = \left|U_{FB}/U\right| <1$, where the
mean-field results are expected to be more accurate.

\begin{figure}
\centerline{\psfig{file=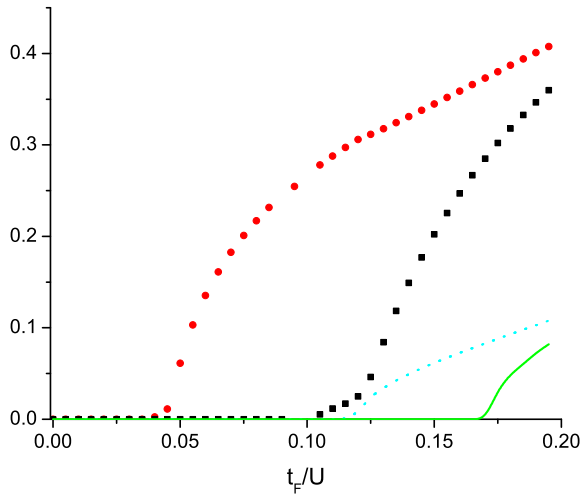,width=\linewidth,angle=0}}
\caption{(Color online) Plot of $\Delta$ and $Z$ for $\mu/U=0.4$,
$\eta=4$ and $\lambda=0$ and $\lambda=0.3$. All symbols have the
same meaning as Fig.\ \ref{fig4}. The plot serves as an illustration
that repulsive interaction between bosons and fermions can enhance
metal/superfluid transitions. } \label{fig6}
\end{figure}
\noindent

To demonstrate the effect of interspecies interaction, we first
concentrate on a fixed value of $\mu_F=\mu_B = 0.7U $ and
$\eta=t_F/t_B= 5$, and study the onset of metal/superfluid-insulator
transition as a function of $t_F/U$ for a few representative values
of $\lambda$. A plot of $Z$ and $\Delta$ for this case, is shown in
Fig.\ \ref{fig4} while the fermionic and bosonic densities are
plotted in Fig.\ \ref{fig5}. The results of these plots can be
understood as following. Consider fixing $\eta$ and gradually
increasing $t_F$ so that the fermions/bosons moves towards a
metal/superfluid-insulator transition point. As long as the
bosons/fermions are in the Mott state, their densities are pinned to
$n_0^B =1$ or $n_0^F=1$, and hence the fermions/bosons see a fixed
chemical potential $\mu_{F(B)}^{\rm eff}= \mu - \lambda U
n_0^B(n_0^F)$. For our chosen value of $\eta$, the metal insulator
transition occurs before the superfluid-insulator transition of the
bosons at $t_F^c(\lambda) = t_F^c\left(\mu_F^{\rm eff}\right)$.
Notice that $t_F^c(\lambda)$ is a non-monotonic function of
$\lambda$ for a given $\mu$ and $\eta$. Hence the metal-insulator
transition for the fermions can either be enhanced ($\lambda=0.5$)
or hindered ($\lambda=0.1$) due to the Bose-Fermi interaction. Once
the fermions have delocalized, the density of fermions changes with
$t_F$ for a fixed chemical potential $\mu$ as seen in Fig.\
\ref{fig5}. Hence the effective chemical potential seen by the
bosons now becomes a function of both $t_B = t_F/\eta$ and
$\lambda$. Thus by increasing $t_B$, we actually traverse a curve
with a finite slope in the $\mu-t_B$ plane in contrast to the
non-interacting ($\lambda=0$) case. Consequently, the
superfluid-insulator transition of the bosons occur at
$t_B^c(\lambda)$ which can be quite different from
$t_B^c(\lambda=0)$. We note this effect may lead to both enhancement
$\left[t_B^c(\lambda)> t_B^c(\lambda=0)\right]$ or hindrance
$\left[t_B^c(\lambda)<t_B^c(\lambda=0)\right]$ of the
superfluid-insulator transition of the bosons depending on the
chosen values of $\eta$ and $\mu$, as can be seen by comparing
Figs.\ \ref{fig4} and \ref{fig6}. For the choice of $\mu/U=0.7$ and
$\eta=5$, we find that $t_B^c(\lambda)> t_B^c(\lambda=0)$, whereas
the reverse case is realized for $\mu/U=0.4$ and $\eta=4$. In the
latter case, both the metal-insulator transition for the fermions
and the superfluid-insulator transition for the bosons are enhanced
for $\lambda>0$. Hence we conclude that a repulsive Bose-Fermi
interaction can either enhance or hinder the onset of
metal/superfluid-insulator transition of a coupled Bose-Fermi
mixture held at a fixed chemical potential. Notice that this effect
is absent if the boson and the fermion densities are held constant
individually, since in that case, there is no influence of the
fermionic metal-insulator transition on the bosonic
superfluid-insulator transition within the mean-field theory. This
is of course an artifact of the present mean-field approximation.
Clearly, the dynamics of the fermions/bosons should play an
important role in the transition. For example, near the
metal-insulator of the fermions, there will be density fluctuations
which give the bosons a chance to hop to a neighboring site even if
they would remain localized in the absence of such fluctuations. A
treatment of this effect requires analysis of the slave-rotor model
beyond the present mean-field approximation and is outside the scope
of the present study.

\begin{figure}
\centerline{\psfig{file=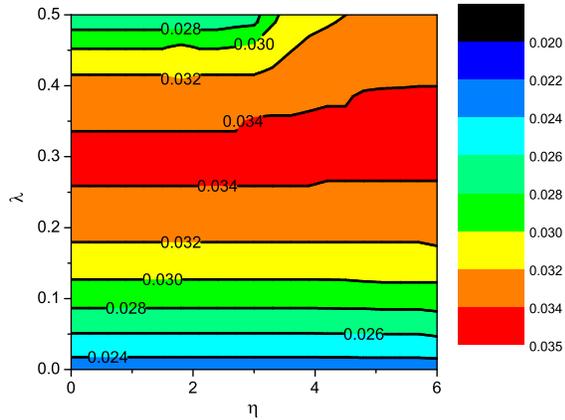,width=\linewidth,angle=0}}
\caption{(Color online) Plot of $t_B^c/U$ as a function of $\eta$
and $\lambda$ for $\mu/U=0.7$. For small $\eta$, the fermions remain
in the Mott state when the superfluid-insulator transition of the
bosons takes place whereas for large $\eta$ they have already
undergone a metal-insulator transition.} \label{fig7}
\end{figure}
\noindent

Finally, we present a plot of the critical hopping strength $t_B^c$
in Fig.\ \ref{fig7} as a function of $\eta$ and $\lambda$ for a
fixed representative $\mu/U=0.7$. We note that at small $\eta$, the
superfluid-insulator transition takes place when the fermions are
still in the Mott state with their density pinned at $n_F^0=1$.
Consequently, the transition for the boson, within the simple
mean-field theory, is the same as that occurring for $\mu_{\rm eff}
= \mu -\lambda U$; $t_B^c$ becomes maximum when $\mu_{\rm eff}/U
\approx 0.4$ or $\lambda \approx 0.3$. However when $\eta$ is large,
the fermions have already undergone the metal-insulator transition
when the boson are at their transition point and hence have $\langle
n_F \rangle \ne 1$. Consequently the bosons see a different
$\mu_{\rm eff} = \mu -\lambda U \langle n_F \rangle$ and hence the
value of $t_B^c$ changes. This is reflected in bending of the phase
boundary in the right half of Fig.\ \ref{fig7}. Analogous plots for
$t_F^c$ will have qualitatively same features.

Before closing this section we would like to make a few qualitative
comments. First, the analysis of the phase diagram for negative
$\lambda$ with  $\left|\lambda\right| \le 1$ can be carried out in a
similar manner and one obtains qualitatively similar results in this
case. Second, at large and negative $\lambda$, the Mott states
correspond to two fermions and $n_0(\lambda)$ bosons localized per
site. In this case, with decreasing lattice potential, the bosons
would undergo a superfluid-insulator transition with the fermions
still in the Mott phase. Therefore the superfluid-insulator
transition of the bosons can be described, within mean-field theory,
by a Hubbard model of bosons with effective chemical potential
$\mu_{\rm eff} = \mu - 2\lambda U $. As we decrease the lattice
depth, the fermions would eventually delocalize when the hopping
coefficient $t_F$ becomes comparable to the energy gap between the
single particle levels ($E_n \simeq 5E_R$) in the potential wells.
However, this is a large energy scale ($E_n \gg U,\, \lambda U$) for
deep lattices and so for any reasonable value of $\eta$, one expects
the superfluid-insulator transition of bosons to occur while the
fermions are still the Mott state. Third, we would like to note that
the slave-rotor mean-field theory worked out here can not be applied
for states with broken translational symmetries in terms of boson
and fermion numbers ( such as the state C shown in Fig.\ \ref{fig3})
since we have used a single-site approximation for the slave-rotor
mean-field theory. Also the present slave-rotor treatment is
expected to be more inaccurate for states with $\lambda \gg 1$ since
the fermion-boson interaction is treated at a mean-field level
within our scheme. Nevertheless we note that qualitatively we would
expect a Mott-superfluid/metal transition from these phases as
$\lambda$ is reduced keeping the density of bosons and fermions
constant. Finally, the present theory can not address the question
of possible superconducting instability of the fermions in the
metallic state. This issue is discussed in Ref.\ \onlinecite{illu1}.
A generalization of our mean-field description of the
superfluid/metal-insulator transitions to address such issues and
inclusion of the effect of quantum fluctuations beyond the
mean-field treatment used here remain open problems to be addressed
in future works.

\section{Experiments}
\label{expt}

A large part of the Mott phases and the superfluid-insulator
transition of the Bose-Fermi mixtures discussed here can be
experimentally accessed using standard experiments on specific
Bose-Fermi mixtures. Before going into details of specific systems,
let us first outline some typical experiments that can be performed
on these systems. One such experiment that is routinely carried out
in atomic systems is measurement of momentum distribution of atoms
in the trap \cite{Greiner1}. This is typically done in a time of
flight measurement by letting the atoms fly out by dropping both the
lattice and the trap and then measuring the position distribution of
the expanding atom cloud. Such a position distribution measurement
yields information about momentum distribution of the atoms within
the trap. These experiments obtain qualitatively different
signatures for atoms in the Mott and the superfluid states for
bosons \cite{Greiner1,Orzel1}, but can not distinguish between the
Bose and the Fermi atoms. To achieve this distinction, one needs to
pass the expanding atom cloud through a Stern-Gerlach apparatus.
Since the fermions and the bosons have different spins (or total
angular momenta), they will get separated during this process and
can thus be distinguished. Such Stern-Gerlach experiments have been
carried out with bosonic atoms in Refs.\ \onlinecite{steg1} and
\onlinecite{wid1} and their generalization to present systems should
be straightforward. Finally, the antiferromagnetic order of fermions
in the Mott phases (either the (1,1) phase in Fig.\ \ref{fig1} or
the phases A and C shown in Fig.\ \ref{fig3}) can be determined by
measuring spatial noise correlations of the expanding cloud in a
time of flight measurement\cite{altman1, greiner2}.

Let us now consider some specific Bose-Fermi mixtures that have been
realized experimentally. One such system is $^6{\rm Li}$ and $^7
{\rm Li}$ mixtures with $a_{FF} = a_{BB}= 5a_0$ and $a_{BF}>0$
\cite{truscott1,schrek1}, where $a_0$ is the Bohr radius. The value
of $a_{BF}$ has not been unambiguously measured in this system, but
is expected to be positive indicating a repulsive interaction
\cite{truscott1}. For this mixture, $m_F \simeq m_B$ and depending
on the value of $a_{BF}$ and by varying the frequencies of the laser
providing the optical lattice (as discussed in Sec.\ \ref{mottsec}),
we may realize different points on the phase diagram shown in Fig.\
\ref{fig1}. Of particular interest is the case where $a_{BF} >
a_{FF}$ \cite{illu1}. In this case, for one atom per site, we expect
to obtain the one of the Mott phases A, B or C shown in Fig.\
\ref{fig3} depending on the specific value of $\lambda \sim
a_{BF}/a_{FF}$ and $\eta=t_B/t_F$ which can be varied by slightly
changing the ratio $V_{F}/V_{B}$ (Eq.\ \ref{param}) \cite{comment1}.
In this case, one can scan a large part of the phase diagram shown
in Fig.\ \ref{fig3}. The possible antiferromagnetic orders of
fermions in phases A or C (Fig.\ \ref{fig3}) can be determined by
measuring the spatial noise correlation measurements \cite{altman1}.
The states B and C can also be distinguished by passing the
expanding clouds through a Stern Gerlach apparatus
\cite{steg1,wid1}. Other possible Mott phases, as shown in Fig.\
\ref{fig1}, are also possible if there are more than one atom per
site. These phases can be detected analogously.

On the other hand, if $a_{BF} < a_{FF}$, one may access the $(1,1)$
phase (Fig.\ \ref{fig1}) with one fermion and one boson per site.
Here in the Mott state, we would find two separate clouds for bosons
and fermions in the Stern-Gerlach measurement with an
antiferromagnetic state for the fermions which can be deduced in the
spatial noise correlation measurement. Further, one can now access
the superfluid/metal-insulator transition in this system by reducing
the depth of the laser producing the optical lattice. The
superfluid-insulator transition can be directly accessed by
measuring the measurement momentum distribution of the bosons. This
provides us a direct measurement of $t_B^c$. One can now also change
$\eta$ as discussed in the last paragraph and access $t_B^c(\eta)$
for a given $\lambda$. This would provide access to a line of
constant $\lambda$ in the phase diagram of Fig. \ref{fig7}.

Another Fermi-Bose mixture which has been realized so far $^{40}
{\rm K}$-$^{87}{\rm Rb}$ mixture \cite{roati1}. Here one expects
$a_{FF} \simeq a_{BB}$ \cite{illu1} and negative $a_{FB}$ indicating
an attractive interaction between bosons and fermions. The magnitude
of $a_{FB}$ is also measured in Ref.\ \onlinecite{roati1} and is
found to be around $3.6 a_{FF}$, although with a large (about
$40\%$) uncertainty. In this system, we expect to find Mott phases
where two fermionic atoms sits on the same site with $n_0(\lambda)$
bosonic atoms. Such states can also be detected in experiments by
passing the expanding clouds through a Stern Gerlach apparatus as
discussed earlier. The superfluid-insulator transition of the bosons
can also be accessed by lowering the lattice depth.

In conclusion, we have studied a Bose-Fermi mixture in an optical
lattice trapped by an optical trap. We have sketched a generic phase
diagram for the possible Mott states of these systems and also
studied the superfluid/metal-insulator transition using a
slave-rotor mean-field theory. We have also discussed definite
experiments that can be performed on specific experimentally
realized systems that can probe at least part of the above-mentioned
phase diagrams.

PM's visit to the Cavendish Lab is supported by the EPSRC (UK) and
by a Visiting Scholars Grant from Trinity College.

\appendix

\section {Comparison of phase diagrams}
\label{appa}

In this section, we compare the phase diagram for a single species
Mott-Hubbard system obtained using the projection operator with
those obtained using mean-field theories \cite{Fisher89,sesh1,van1}
and strong-coupling expansions \cite{freericks1}. To do this, we
consider the case of zero coupling between the bosons and fermions
($\lambda=0$). Our starting point is the Bose-Hubbard Hamiltonian of
Eq.\ \ref{bh} with $\lambda=0$. By tracing the same set of steps, as
in Sec.\ \ref{formalism}, we then obtain the effective Hamiltonian
$H^{\ast}$ (Eq.\ \ref{effham2}). The next step is to obtain the
variational energy using the wavefunction $\Psi_v$ (Eq.\
\ref{varwave}). For the purpose of variational energy computations,
it is sufficient to consider $\Psi_v$ with real coefficients $a$,$b$
and $c$ (Eq.\ \ref{varwave}). This amounts to setting the phase of
the superfluid order parameter $\Delta$ (Eq.\ \ref{ordf}) to zero
and does not affect the variational energy. A straightforward
calculation then yields
\begin{eqnarray}
E_v &=& \left<\Psi_v \right| H^{\ast} \left| \Psi_v\right> = E_0 + E_1 + E_2 \label{varen} \\
E_0 &=&  \delta E_p b^2 + \delta E_h c^2 \label{zeor} \\
E_1 &=& -zt_B a^2 \left[ (n_0 +1) b^2 + n_0 c^2 \right] \label{fior} \\
E_2 &=& -\frac{zt_B^2 n_0\left(n_0+1\right)}{U} \left( a^4 -2b^2c^2\right) \nonumber\\
&& -\frac{z(z-1)t_B^2 n_0 (n_0+1)}{U} \left( a^4(b^2+c^2) -4 a^2 b^2 c^2 \right) \nonumber\\
&& + \frac{2z(z-1)t_B^2\sqrt{n_0+1}}{U} \left[ 2 b c \left(b^2(n_0+1) + c^2 n_0 \right) \right. \nonumber\\
&& \left. - (2n_0 +1) b^2 c^2 \right]a^2 \label{twor}
\end{eqnarray}
where $\delta E_p = \left(-\mu + U n_0\right)$ and $\delta E_h =
\left(\mu - U (n_0-1)\right)$ are the on-site energy costs of adding
a particle and a hole respectively to the Mott phase, and $z=2d$
denotes the coordination number for a $d$ dimensional hypercubic
lattice. The phase diagram can now be obtained by minimizing the
variational energy $E_v$ for given $(t_B/U, \mu/U)$. The
Mott-superfluid phase boundary then corresponds to the minimum value
of $t_c(\mu)$ for which the superfluid order parameter $\Delta$
(Eq.\ \ref{ordf}) is non-zero. Notice that ignoring the {\rm
O}($t_B^2/U^2$) terms amount to setting $E_2=0$. Our numerical
results in this section, however, retains all terms in $E_2$.

Next, we obtain the expression of $t_c^{\rm mf}$ using mean-field
theory. This can be done in a standard manner as shown in Refs.\
\onlinecite{Fisher89,sesh1,van1,subir1} and the mean-field critical
hopping strength in $d$ dimensions can be obtained \cite{subir1}
\begin{eqnarray}
t_c^{\rm mf} = \left(\frac{n_0+1}{U n_0 -\mu} + \frac{n_0}{\mu
-U(n_0-1)} \right)^{-1} . \label{mf}
\end{eqnarray}
where $n_0$ denotes the boson occupation number of the Mott phase
from which the phase boundary is approached.

Finally, we compare the phase diagram obtained from minimizing the
variational energy $E_v$ with the strong-coupling expansion
developed in Ref.\ \onlinecite{freericks1}. The main idea behind the
strong-coupling expansion is that at the phase transition point, for
a given $\mu/U$, the defect state, which corresponds to an
additional particle or hole added to the Mott state, becomes
energetically more favorable. Thus, the energy difference of the
particle or hole defect states with the Mott state given to {\rm
O}($t_B^2/U^2$) by \cite{comment2}
\begin{eqnarray}
\epsilon_p &=& \delta E_p - zt_B (n_0 +1) + \frac{zt_B^2}{2U}
\left(5n_0 +4\right)n_0
\nonumber\\
&& -\frac{z^2t_B^2}{U} n_0 \left(n_0+1\right) \label{defectp}\\
\epsilon_h &=& \delta E_h - zt_B n_0 + \frac{zt_B^2}{2U} \left(5n_0
+1\right)\left(n_0+1\right)
\nonumber\\
&&-\frac{z^2t_B^2}{U} n_0 \left(n_0+1\right) \label{defecth}
\end{eqnarray}
vanishes at $t_B=t_c^{p}\,{\rm or}\, t_c^h$. The phase boundary is
obtained by finding the critical hopping $t_c = {\rm Min}[t_c^p,
t_c^h]$.

Before resorting to numerical evaluation of the phase diagram from
all different techniques, we would like to clarify the following
points. First, although both the defect state calculations of Ref.\
\onlinecite{freericks1} and the projection operator technique
outlined here captures some contributions of $t_B^2/U^2$
fluctuations, they are not identical to each other. To see this, we
consider a second order virtual process for a defect state with one
additional particle(hole) $\left|n_0+(-)1\right>_i
\left|n_0\right>_{j} \, \rightarrow \left|n+(-)2\right>_i
\left|n-(+)1\right>_{j} \, \rightarrow \left|n+(-)2\right>_i
\left|n\right>_{j}$, where $i$ and $j$ are nearest neighbor sites on
the square lattice. This process, which after summing over all
sites, gives ${\rm O}(t_B^2/U^2)$ energy contributions $E_p = -z^2
t_B^2 n_0(n_0+2)/2U$ for particles and $E_h =-z^2 t_B^2
(n_0^2-1)/2U$ for holes. Note that all the states involved in such a
process lie outside the low energy manifold and hence are not
captured within the projection operator technique even if states
with $|n_0\pm 2>$ are incorporated in the variational wavefunction
(Eq.\ \ref{varwave}). On the other hand, the projections operator
technique together with the variational wavefunctions leads to terms
in $E_2$ (Eq.\ \ref{twor}) which involves product of states with one
additional particle and hole (terms which involve product of the
coefficients $b$ and $c$). These terms, which becomes important
mostly near the tip of the Mott lobe, are necessarily absent in
defect state calculations in Ref.\ \onlinecite{freericks1} which
considers energy lowering due to a single particle or hole added
over the Mott state. Thus we find that the best way to compare
different approaches is to compare the phase diagrams obtained using
them.

Second, if we neglect the ${\rm O}(t_B^2/U^2)$ terms, the saddle
point equations for $b$ and $c$ can be easily obtained by minimizing
the variational energy in Eqs.\ \ref{zeor} and \ref{fior}:
\begin{eqnarray}
\left[(n_0+1)-\frac{\delta E_p}{zt_B} \right] b &=&
2(n_0+1)b^3 + n_0 c^2 b \nonumber\\
\left[n_0 - \frac{E_h}{zt_B}\right] c &=& n_0 c^3+ (n_0+1) b^2 c
\label{anmfeq}
\end{eqnarray}
where we have used the constraint $a^2+b^2+c^2=1$. For the Mott
phase, the solution to Eq.\ \ref{anmfeq} is $b=0=c$ and $a=1$. Note
that this ensures that the density in the Mott state is pinned to
$\left<n\right> = n_0$. Solutions with non-zero $b$ and $c$ occurs
when in the SF phase for which $t_B \ge t_c$. We find that in the
superfluid phase near the Mott transition line, Eq.\ \ref{anmfeq}
admits the following solutions. For $\mu/U \le n_0/(2n_0+1) $ one
gets $t_c = t_c^{h}= \delta E_h/zn_0$ and
\begin{eqnarray}
c&=& \frac{1}{\sqrt{2}} \left(1-\frac{t_c^h}{t_B}\right)^{1/2} \quad
b=0, \label{sol1}
\end{eqnarray}
whereas for $\mu/U \ge n_0/(2n_0+1)$ one has $t_c = t_c^p= \delta
E_p/z(n_0+1)$ and
\begin{eqnarray}
b &=& \frac{1}{\sqrt{2}} \left(1-\frac{t_c^p}{t_B}\right)^{1/2},
\quad c=0. \label{sol2}
\end{eqnarray}
Eqs. \ref{sol1} and \ref{sol2} shows that near the phase transition
$b,c \ll a$, which is crucial to our analysis. Further, as normally
expected in a second order quantum phase transition, the
coefficients $b$ and $c$ are continuous across the transition. Next,
we discuss inclusion of ${\rm O}(t_B^2/U^2)$ terms. If these terms
(Eq.\ \ref{twor}) are included, obtaining analytical solutions for b
and c becomes difficult as it amounts to solving two coupled cubic
equations. However, we have checked during numerical evaluation of
the phase diagram minimizing Eq.\ \ref{varen} ( which retains second
order terms in Eq.\ \ref{twor}) that $b$ and $c$ are always small
compared to $a$ near the phase transition and that near the ends of
the Mott lobe their numerical values are well reproduced by Eqs.\
\ref{sol1} and \ref{sol2}.

Finally, we comment about our choice of variational wavefunction in
which we have only retained states with one additional particle and
hole per site over the parent Mott state. In principle, one can
retain states with two or more particle per sites which leads to a
more complicated trial wave function. For example, one can retain
the states $\left|n_0 \pm 2\right>$ so that the variational
wavefunctions become
\begin{eqnarray}
\left|\psi'\right> &=& \prod_i \left|\psi'\right>_i \nonumber\\
\left|\psi'\right>_i &=& a\left|n_0 \right> + b\left|n_0 +1 \right>
+c \left|n_0 -1 \right> \nonumber\\
&& + d \left|n_0 +2  \right> + e \left|n_0 - 2 \right> \label{var2a}
\end{eqnarray}
which leads to a variational energy to ${\rm O}(t_B/U)$ (Eq.\
\ref{projham}) $E'_{v} = \left<\psi'\right| H_0(\lambda=0) + T^0
\left|\psi'\right> = E'_{0v}+ E'_{1v}$, where,
\begin{eqnarray}
E'_{0v} &=& \delta E_p b^2 + \delta E_h c^2 + \delta E_{2p} d^2 +
\delta E_{2h} e^2 \nonumber\\
\delta E_{2p}&=& -2\mu + U(2n_0+1) \quad \delta E_{2h} = 2 \mu
-3U(n_0-1) \nonumber\\
E'_{1v} &=& -zt_b \Bigg[(n_0+1)a^2b^2 + n_0 a^2 c^2 + (n_0-1) c^2
e^2
 \nonumber\\
&& + \sqrt{n_0(n_0-1)} c^2 a e + \sqrt{(n_0+2)(n_0+1)}b^2 a d
\nonumber\\
&&+ 2 \sqrt{(n_0-1)(n_0+2)} b c d e +  (n_0+2) d^2 b^2\nonumber\\
&&+  abc \left(d \sqrt{n_0(n_0+2)} + e \sqrt{n_0^2-1} \right) \Bigg]
\label{var2b}
\end{eqnarray}
To analyze the phase diagram obtained from Eqs.\ \ref{var2a} and
\ref{var2b}, we first note that in the Mott phase $a=1$ and
$b=c=d=e=0$, as discussed before. Next let us discuss the
minimization of $E'_{v}$ near the Mott-SF transition line. When
$d=e=0$, the MI-SF transition occurs at $zt_c = {\rm
Min}(t_c^p,t_c^h)$, with $b,c \ll 1$ and $a \simeq 1$. So to find
out whether non-zero $d$ and/or $e$ is favorable for energy
minimization, we need to find the effect of turning on a non-zero
$d$ and/or $e$ near this transition point and check whether it leads
to lowering of the variational energy. We find that turning on a
non-zero $d$ and/or $e$ near this point, as can be seen from the
expression of $E'_{1v}$ (Eq.\ \ref{var2b}), leads to energy gain $
\ll zt_c$ (since in the expression of $E'_{1v}$, $d$ and $e$ always
appear as a product with $b$ and $c$ which are small), whereas the
energy cost (Eq.\ \ref{var2a}) is ${\rm O}(\delta E_{2p})$ and/or
${\rm O}(\delta E_{2h})$. It turns out that the latter is always
greater than the former which leads to a net non-zero energy cost.
As a specific example to demonstrate this point, let us consider
$\mu \simeq U$ so that for $t_B\simeq t_c = t_c^p$, $b\ne 0$ and
$c=0$ (Eq.\ \ref{sol2}). Then turning on a non-zero $d$ results in
an energy change $\delta E = d \left(\delta E_{2p}d -zt_c
\left[(n_0+2)b^2 d +\sqrt{(n_0+2)(n_0+1)}b^2 a\right] \right) > 0$
since $b,d \ll 1$ and $\delta E_{2h} -zt_c \simeq U(n_0+1)-\mu >0$.
Thus near $t_B=t_c$, the energy cost from $E'_{0v}$ clearly
outweighs energy gain from $E'_{1v}$ (since $zt_c/U \ll 1$ for all
parameter regime and $b,c,d,e \ll 1$ near $t_B=t_c$) and hence the
coefficients $d$ and $e$ remain vanishingly small at the phase
transition. As we move inside the superfluid phase and $t_B$ becomes
large compared to $t_c$, these coefficients becomes significant. In
the present work, we have always restricted ourselves to regions
sufficiently near the transition line where the coefficients $d$ and
$e$ are small. We note that we have explicitly checked numerically
from the full variational energy ( including ${\rm O}(t_B^2/U^2)$
terms which we have not written down explicitly here to avoid
clutter), that the above-mentioned qualitative argument always holds
for all the regimes studied in the present work. We shall therefore
neglect these terms in the rest of this section.

\begin{figure}
\centerline{\psfig{file=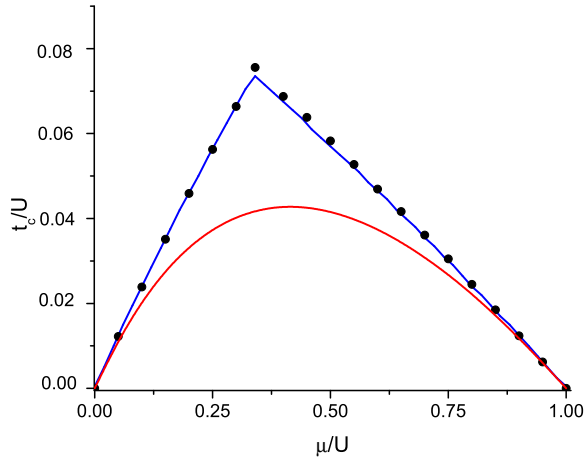,width=\linewidth,angle=0}}
\caption{(Color online) Phase diagram for $d=2$ and $n_0=1$
calculated by minimizing variational energy $E_v$ (black dots),
mean-field theory (red solid line) and defect energy calculation to
{\rm O}($t_B^2/U^2$) (blue solid line). The phase boundaries
computed by minimizing $E_v$ and by defect energy calculations
compare well to each other but differs from the mean-field theory
near the tip of the Mott lobe. } \label{fig8}
\end{figure}
\noindent

\begin{figure}
\centerline{\psfig{file=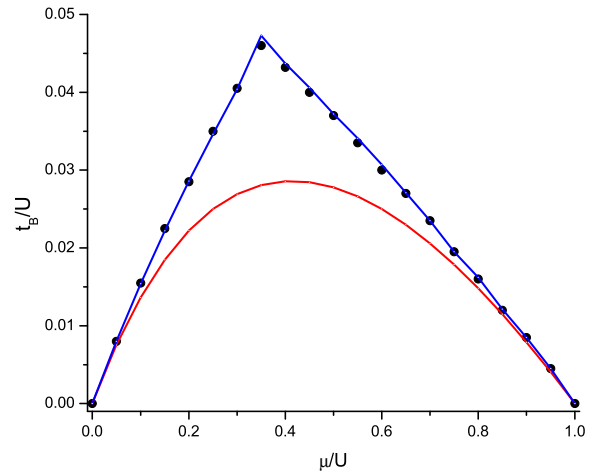,width=\linewidth,angle=0}}
\caption{(Color online) Phase diagram for $d=3$ and $n_0=1$. All
notations are same in Fig.\ \ref{fig8}. As in 2D, the phase
boundaries computed by minimizing $E_v$ and by defect energy
calculations compare well to each other but differs from the
mean-field theory near the tip of the Mott lobe.} \label{fig10}
\end{figure}

To compare the results from different approaches, we now plot the
phase diagrams for $n_0=1$ obtained by minimizing the full
variational energy $E_v$ to ${\rm O}(t_B^2/U^2)$(Eq.\ \ref{varen}),
from the mean-field equation (Eq.\ \ref{mf}), and by computing the
energy of the defect states to {\rm O}($t_B^2/U^2$)
\cite{freericks1}. These plots are shown in Fig.\ \ref{fig8} for
$d=2$ and in Fig.\ \ref{fig10} for $d=3$. We find that in spite of
the dissimilarity of the two approaches discussed above, the
numerical phase boundary obtained by minimizing $E_v$ matches that
obtained from {\rm O}($t_B^2/U^2$) defect state calculation of Ref.\
\onlinecite{freericks1} quite well in both cases, but differs
substantially from the phase boundary obtained using the mean-field
theory. Further, for $d=2$, Quantum Monte Carlo data, available for
the critical hopping at the tip of the Mott lobe
\cite{freericks1,nandini1}, predicts $(t_c/U)_{\rm MC} =0.061 \pm
0.006$. The corresponding values obtained from minimization of $E_v$
and {\rm O}($t_B^2/U^2$) defect state energy calculation of Ref.\
\onlinecite{freericks1} are $(t_c/U)_{\rm var} =0.0755$ and
$(t_c/U)_{\rm defect} =0.0735$ respectively. The mean-field theory
predicts a value $(t_c/U)_{\rm mf}= 0.041$ while an {\rm
O}($t_B^3/U^3$) calculation of defect states energies
\cite{freericks1} gives $(t_c/U)_{\rm defect} =0.068$. Thus we
conclude that the phase boundary obtained from our variational
energy calculation scheme compares well with the defect state energy
calculation to {\rm O}($t_B^2/U^2$).

\noindent

In $d=3$, we can compare our results in Fig.\ \ref{fig10} to Figs.\
\ref{fig4}, \ref{fig5} and \ref{fig6} for $\lambda=0$ and $\mu=0.7$
and $\mu=0.4$ respectively. We find that the phase diagram obtained
using the projection operator techniques predicts $t_c/U=0.0232$ for
$\mu/U=0.7$ and $\eta=5$ (Fig.\ \ref{fig4})  and predicts
$t_c/U=0.0431$ for $\mu/U=0.4$ and $\eta=4$, (Fig.\ \ref{fig6}). As
seen from Fig. \ref{fig10}, these values compares well to those
obtained from defect state calculations to $O(t_B^2/U^2)$. However,
they do not compare favorably with $t_c^{\rm mf}$ which predicts
$t_c^{\rm mf}/U=0.0208$ for $\mu/U=0.7$ and $t_c^{\rm mf}/U=0.0266$
for $\mu/U=0.4$. Further, the corresponding values of $t_c/U$ from
the third order defect state calculation are $0.0225$ for
$\mu/U=0.7$ and $0.0405$ for $\mu/U=0.4$ which compares quite
favorably to the projection operator method, but not to the
mean-field theory. At the tip of the Mott lobe ($\mu/U \simeq
0.37$), where the difference between results obtained from different
methods become most apparent,  the values of $t_c/U$ obtained from
different methods are $0.0419$ (third order defect state), $0.0459$
(second order defect state), $0.0451$ (projection operator
technique), $0.0276$ (mean field) and $0.03480( 2)$ (very recently
available Quantum Monte Carlo results \cite{svref}).


\begin{thebibliography}{99}

\bibitem{Greiner1} M. Greiner, O. Mandel, T. Esslinger, T.W. Ha¨nsch, and I. Bloch,
Nature(London) {\bf 415}, 39 (2002); M. Greiner, O. Mandel, T.W.
Hänsch, and I. Bloch, Nature {\bf 419}, 51 (2003)

\bibitem{Orzel1} C. Orzel, A.K. Tuchman, M.L. Fenselau, M. Yasuda, and M.A.
Kasevich, Science {\bf 291}, 2386 (2001).

\bibitem{Fisher89} M. P. A. Fisher, P. B. Weichman, G. Grinstein, and D. S.
Fisher, Phys. Rev. B {\bf 40}, 546 (1989);

\bibitem{Jaksch1} D. Jaksch, C. Bruder, J. I. Cirac, C. W. Gardiner, and P. Zoller,
Phys. Rev. Lett. {\bf 81}, 3108 (1998);

\bibitem{sesh1} K. Sheshadri, H. R. Krishnamurthy, R. Pandit, and T. V.
Ramakrishnan, Europhys. Lett. {\bf 22}, 257 (1993)

\bibitem{van1} D. Van Oosten, P. van der Straten, and H. T. C. Stoof, Phys. Rev.
A {\bf 63}, 053601 (2001).

\bibitem {Sengupta1} K. Sengupta and N. Dupuis, Phys. Rev. A {\bf 71}, 033629
(2005).

\bibitem {imambekov1} A. Imambekov, M. Lukin, and E. Demler, Phys. Rev. A {\bf 68},
063602 (2003).

\bibitem {prokofiev1} A. Kuklov and B. Svistunov, Phys. Rev. Lett. {\bf 90}, 100401
(2003); A. Kuklov, N. Prokof'ev, and B. Svistunov, Phys. Rev. Lett.
{\bf 92}, 050402 (2004).

\bibitem{demler1} L-M. Duan, E. Demler, and M. Lukin, Phys. Rev. Lett.
{\bf 91}, 090402 (2003).

\bibitem{issacson1} A. Isacsson, Min-Chul Cha, K. Sengupta, and S.M.
Girvin, Phys. Rev. B {\bf 72}, 184507 (2005).

\bibitem{Greiner2} M. Greiner, C. A. Regal, and D.S. Jin, Nature {\bf 426}, 537-540
(2003).

\bibitem{Zwierlein1} M.W. Zwierlein, J.R. Abo-Shaeer, A. Schirotzek,
and W. Ketterle, Nature {\bf 435}, 1047 (2005).

\bibitem{exprev1} For a review, see R. Onofrio and C. Presilla, J.
Stat. Phys. {\bf 115} 57 (2004).

\bibitem{modugno1} G. Roati, E. de Mirandes, F. Ferlaino, H. Ott, G. Modugno and
M. Inguscio, Phys. Rev. Lett. {\bf 92}, 230402 (2004); G. Modugno,
E. de Mirandes, F. Ferlando, H. Ott, G. Roati, and M. Inguscio, AIP
Conf. Proc. {\bf 770}, 197 (2005).

\bibitem{roth1} R. Roth and K. Burnett, Phys. Rev. A {\bf 69}, 021601
(2004).

\bibitem{albus1} A. Albus, F. Illumunati, and J. Eisert, Phys. Rev. A {\bf 68},
023606 (2003)

\bibitem{lewenstein1} M. Lewenstein, L. Santos, M.A. Baranov, and H.
Fehrmann, Phys. Rev. Lett. {\bf 92}, 050401 (2004).

\bibitem{cramer1} M. Cramer, J. Eisert, and F. Illuminati, Phys. Rev. Lett.
{\bf 93}, 190405 (2004).

\bibitem{yu1} Y. Yu and S.T. Chui, Phys. Rev. A {\bf 71}, 033608
(2005).


\bibitem{illu1} F. Illumunati and A. Albus, Phys. Rev. Lett. {\bf 93}, 090406
(2004).

\bibitem{carr1} L.D. Carr and M. Holland, Phys. Rev. A {\bf 72}, 031604
(2005).

\bibitem{comment0} More precisely, this means that the Fermions in
the mixture is of the same species but can belong to two different
total angular momentum (or generalized spin) states \cite{illu1}.

\bibitem{florens1} S. Florens and A. Georges, Phys. Rev. B {\bf 70}, 035114
(2004);  A. Georges, S. Florens, and T.A. Costi, J. Physique IV {\bf
114}, 165 (2004).

\bibitem{florian1} C. Schroll, F. Marquadt, and C. Bruder, Phys. Rev. A
{\bf 70}, 053609 (2004).

\bibitem{freericks1} J.K. Freericks and H. Monien, Europhys. Lett. {\bf 26}
545 (1994); {\it ibid} Phys. Rev. B {\bf 53} 2691 (1996).

\bibitem{steg1} J. Stegner {\it et al.}, Nature {\bf 396}, 345
(1998).

\bibitem{wid1} J. Widera {\it et al.}, cond-mat/0505492
(unpublished).

\bibitem{altman1} E. Altman, E. Demler, and M.D. Lukin, Phys. Rev. A {\bf 70},
013603 (2004).

\bibitem{greiner2} M. Greiner, C. A. Regal, J. T. Stewart, and D. S. Jin,
Phys. Rev. Lett. {\bf 94}, 110401 (2005)

\bibitem{truscott1} A.G. Truscott {\it et al.}, Science {\bf 291},
2570 (2004).

\bibitem{schrek1} F. Schrek {\it et al.}, Phys. Rev. Lett. {\bf 87},
080403 (2001).

\bibitem{comment1} Such a variation also changes $U_{FF}/U_{BB}$ in
principle. But one can achieve an appreciable change in $\eta$ by
maintaining a small change in $U_{FF}/U_{BB}$ since $t_F$ has an
exponetial dependence on the lattice potential.

\bibitem{roati1} G. Roati, F. Riboli, G. Modugno, and M. Inguscio, Phys. Rev. Lett.
{\bf 89}, 150403 (2002); Ferlaino {it et\, al.}, Phys. rev. A {\bf
73}, 040702 (2006).

\bibitem{comment2} The defect state energies are calculated to {\rm O}($t_B^3/U^3$) in
Ref.\ \onlinecite{freericks1}. Here we have only retained terms to
{\rm O}($t_B^2/U^2$) for the purpose of comparison with our approach
in Fig.\ \ref{fig8} and \ref{fig10}.

\bibitem{nandini1} W. Krauth and N. Trivedi, EuroPhys. Lett. {\bf 14}, 627 (1991);
W. Krauth, N. Trivedi and D. Ceperley, Phys. Rev. Lett. {\bf 67},
2307 (1991).

\bibitem{subir1} S. Sachdev, {\it Quantum Phase Transitions},
Cambridge University Press, Cambridge, UK (1999).

\bibitem{svref} B. Caprogrosso-Sansone, N. Prokof'ev, and B.V.
Svistunov, cond-mat/0701178 (unpublished).

\end{thebibliography}
\end{document}